\documentclass[12pt]{article}
\usepackage{epsfig,amssymb,amsmath,psfrag}

\catcode`\@=11
\font\cmss=cmss12 
\def\1{\hbox{{1}\kern-.25em\hbox{l}}}
\def\bfZ{\relax{\hbox{\cmss Z\kern-.4em Z}}}

\def \be  {\begin{equation}}
\def \ee  {\end{equation}}
\def \ba  {\begin{eqnarray}}
\def \ea  {\end{eqnarray}}
\def \baa {\begin{eqnarray*}}
\def \eaa {\end{eqnarray*}}
\def \bb  {\begin {thebibliography} }
\def \eb  {\end{thebibliography}}
\def \lab #1 {\label{#1}}

\newcommand\re[1]{(\ref{#1})}

\def \matrix #1 {\left(\begin{array}{cc} #1 \end{array}\right)}

\newcommand{\as}{\ifmmode\alpha_{\rm s}\else{$\alpha_{\rm s}$}\fi}
\newcommand{\asbar}{\ifmmode\bar{\alpha}_{\rm s}\else{$\bar{\alpha}_{\rm s}$}\fi}

\newcommand{\ft}[2]{{\textstyle\frac{#1}{#2}}}

\font\cmss=cmss12 
\def\inbar{\,\vrule height1.5ex width.4pt depth0pt}
\def\IC{\relax\hbox{$\inbar\kern-.3em{\rm C}$}}
\def\IZ{\relax{\hbox{\cmss Z\kern-.4em Z}}}
\def\IR{{\hbox{{\rm I}\kern-.2em\hbox{\rm R}}}}

\def\IP{{\hbox{{\rm I}\kern-.2em\hbox{\rm P}}}}
\def\II{\hbox{{1}\kern-.25em\hbox{l}}}

\def\numberbysection{\@addtoreset{equation}{section}
                     \def\theequation{\thesection.\arabic{equation}}}
\numberbysection

\newbox\lett\newdimen\lheight\newdimen\lwidth
\def\ontop#1#2{\setbox\lett=\hbox{#2}\lheight\ht\lett
\multiply\lheight by 12 \divide\lheight by 10\relax%
\lwidth\wd\lett \multiply\lwidth by 8 \divide\lwidth by 10\relax #2\kern-\lwidth%
\raise\lheight\hbox{{$\scriptstyle #1$}}\kern.1ex}

\def\XXint#1#2#3{{\setbox0=\hbox{$#1{#2#3}{\int}$}
     \vcenter{\hbox{$#2#3$}}\kern-.5\wd0}}

\begin{document}

\begin{titlepage}

\vskip2cm

\centerline{\large \bf Long-range $SL(2)$ Baxter equation
                       in ${\mathcal N} = 4$ super-Yang-Mills theory}

\vspace{1cm}

\centerline{\sc A.V. Belitsky}

\vspace{10mm}

\centerline{\it Department of Physics, Arizona State University}
\centerline{\it Tempe, AZ 85287-1504, USA}

\vspace{1cm}

\centerline{\bf Abstract}

\vspace{5mm}

Relying on a few lowest order perturbative calculations of anomalous dimensions of
gauge invariant operators built from holomorphic scalar fields and an arbitrary
number of covariant derivatives in maximally supersymmetric gauge theory, we propose
an all-loop generalization of the Baxter equation which determines their spectrum.
The equation does not take into account wrapping effects and is thus asymptotic in
character. We develop an asymptotic expansion of the deformed Baxter equation for
large values of the conformal spin and derive an integral equation for the cusp
anomalous dimension.

\end{titlepage}

\setcounter{footnote} 0

\thispagestyle{empty}

\newpage

\pagestyle{plain} \setcounter{page} 1

\section{Introduction}

Four-dimensional nonabelian gauge theories were found to possess integrable structures.
Initially they had been discovered in evolution equations for scattering amplitudes in
multi-color limit of high-energy QCD \cite{Lip93,FadKor95}. Subsequently it was demonstrated
that hidden symmetries also arise in the field-theoretical dilatation operator whose eigenvalues
determine anomalous dimensions of gauge invariant composite operators of elementary fields in
underlying models. Integrability was revealed in one-loop anomalous dimensions of twist-$L$
maximal-helicity quasipartonic Wilson operators in QCD \cite{BukFroKurLip85} by identifying
the former with eigenenergies of the $L-$site XXX Heisenberg spin chain \cite{BraDerMan98}.
The magnet turns out to be noncompact, for the spin operators acting on its sites transform
in the infinite-dimensional representation of the collinear subgroup $SL(2, \mathbb{R})$ of
the conformal group $SO(4,2)$. Since the one-loop phenomenon is spawned by gluons, invariably
present in Yang-Mills theories,---supersymmetric or not,---they all necessarily exhibit the
same, universal integrable structures. The differences arise due to distinct particle
contents of the models: while only holomorphic sectors are integrable in QCD and its nearest
supersymmetric $\mathcal{N} = 1, 2$ siblings \cite{BelDerKorMan04}, the maximal supersymmetry
of the $\mathcal{N} = 4$ super-Yang-Mills theory extends integrability to all operators
\cite{Lip98,BeiKriSta03,Bei04,BelDerKorMan04}. Recent perturbative studies build up a
growing amount of direct evidence that integrability persists in certain closed compact
\cite{BeiKriSta03,Bei04} and noncompact \cite{Ede04,BelKorMul05,Zwi06,EdeSta06,BelKorMul06}
subsectors of gauge theories even in higher orders of perturbation theory. Thus, while ruled
out for gauge theories with $\mathcal{N} < 4$ supercharges, it is plausible that the maximally
supersymmetric Yang-Mills theory is completely integrable. An additional confirmation for this
conjecture comes from studies of multi-loop multi-leg scattering amplitudes which display
intriguing iterative structures \cite{AnaBerDixKos03,BerDixSmi05}. These arguments suggest
that the spectrum of all-loop anomalous dimensions in $\mathcal{N} = 4$ SYM theory is determined
by a putative long-range integrable spin chain with the dilatation operator being its Hamiltonian.

In this note we probe the underlying integrable long-range magnet by proposing its multi-loop
perturbative structure within the framework of the Baxter $\mathbb{Q}-$operator \cite{Bax72}.
This approach is based on the existence of an operator $\mathbb{Q} (u)$ depending on a spectral
parameter $u$ and acting on the Hilbert space of the magnet. For different values of $u$ it
forms a family of mutually commuting operators, simultaneously commuting with the spin-chain
Hamiltonian as well. Although in the present circumstances, the formalism is equivalent to the
Bethe Ansatz approach, it possesses certain advantages. First, the eigenvalue $Q (u)$ of the
Baxter operator $\mathbb{Q} (u)$ determines the single-particle wave function of the chain in
the representation of separated variables \cite{Skl85}. Second, the equation for the
$\mathbb{Q}-$operator,---known as the Baxter equation,---is polynomial, to be contrasted with
a set of coupled transcendental Bethe equations. Third, it allows for a straightforward
asymptotic analysis when quantum numbers of the chain are large as will be demonstrated below.

Currently we restrict our consideration to the closed noncompact $SL(2)$ sector of the gauge
\cite{BelGorKor03,Bei03} theory which is spanned by single-trace maximal $R-$charge Wilson
operators built from the holomorphic scalar fields $X = \phi_1 + i \phi_2$ and covariant
derivatives,
\be
\label{ScalarWilsonOperators}
\mathcal{O}_{n_1 n_2 \dots n_L} (0)
=
{\rm tr}
\left\{
(i D_+)^{n_1} X(0)
(i D_+)^{n_2} X(0)
\dots
(i D_+)^{n_L} X(0)
\right\}
\, .
\ee
Here $D_+ = D_\mu n^\mu$ is projected on the light cone with a null vector $n^\mu$, $n^2 = 0$, in
order to factor out the maximal Lorentz-spin component from the operator in question. These Wilson
operators mix with each other under renormalization group evolution and acquire anomalous dimensions
at all orders of perturbative series in coupling constant\footnote{Their complete two-loop planar
mixing matrix has been recently computed in Ref.\ \cite{BelKorMul06}.}
\be
\gamma (g) = \sum_{n = 1}^\infty g^{2 n} \gamma^{(n)}
\, .
\ee
We find it convenient to use the expansion parameter $g$ related to the 't Hooft coupling constant
$\lambda$ via
\be
g = \sqrt{2 \lambda} = \frac{g_{\rm\scriptscriptstyle YM} \sqrt{N_c}}{2 \pi}
\, .
\ee
The anomalous dimension $\gamma (g)$ depends on parameters characterizing the operator: its twist
$L$, determined by the number of $X-$fields, and its Lorentz spin $N = n_1 + n_2 + \dots + n_L$.
Within the method of the Baxter $\mathbb{Q}-$operator, the eigenspectrum of one-loop anomalous
dimensions $\gamma^{{\scriptscriptstyle (0)}}$ and the corresponding quasimomentum
$\theta^{{\scriptscriptstyle (0)}}$ are determined by the leading order Baxter function
$Q^{{\scriptscriptstyle (0)}} (u)$
\be
\label{LOenergyANDquasimom}
\gamma^{{\scriptscriptstyle (0)}}
=
\ft{i}{2} \left[ \ln Q^{{\scriptscriptstyle (0)}} (\ft{i}{2}) \right]^\prime
-
\ft{i}{2} \left[ \ln Q^{{\scriptscriptstyle (0)}} (- \ft{i}{2}) \right]^\prime
\, , \qquad
\theta^{{\scriptscriptstyle (0)}}
=
\ln Q^{{\scriptscriptstyle (0)}} (\ft{i}{2}) - \ln Q^{{\scriptscriptstyle (0)}} (- \ft{i}{2})
\, .
\ee
Since the Baxter function $Q^{{\scriptscriptstyle (0)}} (u)$ is related to the eigenfunction of
the mixing matrix, it corresponds to a multiplicatively renormalizable Wilson operator and thus
has to be polynomial in $u$ of order $N$, $Q^{{\scriptscriptstyle (0)}} (u) = (u - u^{
{\scriptscriptstyle (0)}}_1) (u - u^{{\scriptscriptstyle (0)}}_2) \dots (u - u^{
{\scriptscriptstyle (0)}}_N)$. The zeros of this polynomial are determined by the Bethe roots
$u^{{\scriptscriptstyle (0)}}_n$ which take only real values for the noncompact $SL
(2, \mathbb{R})$ spin chain \cite{FadKor95,Kor95}. The function $Q^{{\scriptscriptstyle (0)}} (u)$
obeys the finite-difference Baxter equation \cite{Bax72}
\be
\label{OneLoopBaxterEq}
u_+^L Q^{{\scriptscriptstyle (0)}} (u + i)
+
u_-^L Q^{{\scriptscriptstyle (0)}} (u - i)
=
t^{{\scriptscriptstyle (0)}} (u) Q^{{\scriptscriptstyle (0)}} (u)
\, ,
\ee
where the spectral parameter in the dressing factors $u_\pm^L$ is shifted by the conformal spin $s =
\ft12$ of the scalar field $X$, $u_\pm = u \pm \frac{i}{2}$ and $t^{{\scriptscriptstyle (0)}} (u)$ is
an order$-L$ polynomial in $u$ depending on the integrals of motion.

\section{Three-loop Baxter equation}

Explicit perturbative calculations \cite{Ede04,BelKorMul05,EdeSta06} of two-loop corrections to
the anomalous dimensions of the scalar operators \re{ScalarWilsonOperators} exhibit double
degeneracy of energy levels with zero quasimomentum. This hints at the existence of nontrivial
odd-parity conserved charges and thus persistence of integrability at higher orders of perturbation
theory.

Beyond one loop, the formalism of the Baxter operator gets modified accordingly. The Bethe roots
acquire corrections in coupling constant to all orders of perturbation theory,
\be
u_n (g) = \sum_{k = 0}^\infty g^{2 k} u^{(k)}_n
\, ,
\ee
and obey deformed Bethe Ansatz equations \cite{BeiSta05}. The reality of Bethe roots $u_k
(g)$ have to be preserved to all orders since the eigenvalue $Q (u)$ of $\mathbb{Q} (u)$ is a wave
function of the chain with the number of its nodes on the real $u-$axis coinciding with the spin
$N$ of the operator. The polynomial
\be
Q (u) = \prod_{n = 1}^N (u - u_n (g))
\, ,
\ee
fulfills these properties and is real $Q^\ast (u) = Q (u^\ast)$ for $u^\ast = u$. In Ref.\
\cite{BelKorMul06} we found from available two- \cite{KotLip02,KotLip03,Ede04,BelKorMul05}
and three-loop \cite{EdeJarSok04,MocVerVog04,KotLipOniVel04} diagrammatic calculations of
anomalous dimensions that the Baxter equation possesses the form
\be
\label{AllOrderBaxterEq}
x_+^L {\rm e}^{\sigma_+ (x_+)} Q (u + i)
+
x_-^L {\rm e}^{\sigma_- (x_-)} Q (u - i)
=
t (u) Q (u)
\, ,
\ee
with the dressing factors depending on the renormalized spectral parameter \cite{BeiDipSta04}
\be
x [u] = \frac{1}{2} \left( u + \sqrt{u^2 - g^2} \right)
\, , \qquad
x_\pm = x [u_\pm]
\, .
\ee
The multi-loop transfer matrix%
\footnote{Note that with this transfer matrix the resulting Baxter equation breaks down already
at order $\mathcal{O} (g^{2 n})$ in coupling constant with $n = L$. It turns out that one can
correctly incorporate order $n = L$ corrections by replacing the leading term in $t (u)$ with
the following combination $2 u^L \to x^L_+ + x^L_- - (\ft{i}{2})^L - (- \ft{i}{2})^L$.}
\be
t (u) = 2 u^L + q_1 (g) u^{L - 1} + q_2 (g) u^{L - 2} + \dots + q_L (g)
\ee
acquires the ``missing" term $\sim u^{L - 1}$ at $g^2-$order, i.e., $q_1 (g) \sim \mathcal{O}
(g^2)$, while the rest of the charges start from $\mathcal{O} (g^0)$, $q_k (g) =
q_k^{{\scriptscriptstyle (0)}} + \mathcal{O} (g^2)$. The additional dressing factors $\sigma_\pm$
obey the complex conjugation condition $\left( \sigma_- (x_-) \right)^\ast = \sigma_+ (x_+)$
for $\Im{\rm m} u = 0$ and encode the renormalization of the noncompact charges $q_k (g)$ at higher
orders. An analysis yielded the following result to three-loop order \cite{BelKorMul06}
\be
\label{PertExpan-sigma}
\sigma_\pm (x)
=
-
\frac{g^2}{2 x} \left[ \ln Q (\pm \ft{i}{2}) \right]^\prime
-
\frac{g^4}{16 x^2}
\left\{
\left[ \ln Q (\pm \ft{i}{2}) \right]^{\prime\prime}
+
x \left[ \ln Q (\pm \ft{i}{2}) \right]^{\prime\prime\prime}
\right\}
+
\mathcal{O} (g^6)
\, .
\ee
While the anomalous dimension is expressed order-by-order in coupling constant $g$ in terms of the
solution to Eq.\ \re{AllOrderBaxterEq} as \cite{BelKorMul06}
\be
\label{ADs3loops}
\gamma (g)
=
i \left\{
\frac{g^2}{2} \left[ \ln Q (u) \right]^\prime
+
\frac{ g^4}{16} \left[ \ln Q (u) \right]^{\prime\prime\prime}
+
\frac{g^6}{384} \left[ \ln Q (u) \right]^{(5)}
+
\mathcal{O} (g^8)
\right\}^{u = i/2}_{u = - i/2}
\, .
\ee
The anomalous dimensions found using these equations reproduced exactly available perturbative
predictions. One can demonstrate that the condition of the pole-free transfer matrix at Bethe
roots $u_n (g)$, $t (u_n) = 0$ immediately produces the three-loop Bethe Ansatz of Ref.\
\cite{Sta04}.

\section{Multi-loop conjecture}

The above representation \re{PertExpan-sigma} of the dressing factors $\sigma_\pm$ can be brought to
a very suggestive form. Namely, a quick inspection allows one to rewrite these terms as an expansion
in terms of the Chebyshev polynomials of the second kind $U_k$,
\be
\label{DressingInfiniteSeries}
\sigma_\pm (x)
=
\frac{2 g}{\pi} \int_{-1}^1 dt \, \sqrt{1 - t^2}
\left[ \ln Q (\pm \ft{i}{2} - g t) \right]^\prime
\sum_{n = 0}^{n_{\rm max}}
\left( - \frac{g}{2 x} \right)^{n + 1} \frac{U_n (t)}{n + 1}
\, ,
\ee
with $n_{\rm max} = 2$, valid to $\mathcal{O} (g^3)$ in the approximation of Eq.\ \re{PertExpan-sigma}.
Having this representation at our disposal, we may naturally extend the first few terms of the
available perturbative series to all orders in coupling $g$, by sending $n_{\rm max} \to \infty$.
Using the summation theorem for Chebyshev polynomials, one can sum the infinite series up into the
function $- ({\rm arccot} \frac{t + 2 x / g}{\sqrt{1 - t^2}})/\sqrt{1 - t^2}$ and, upon a variable
transformation, write $\sigma_\pm$ in the form (with $\bar{z} = 1 - z$)
\ba
\label{AllLoopSigma}
\sigma_\pm (x)
\!\!\!&=&\!\!\!
- \frac{g^2}{2 \pi x} \int_0^1 dz \int_{-1}^1 \frac{dt}{\sqrt{1 - t^2}}
\left[
\ln Q
\left( \pm \ft{i}{2} - g \sqrt{z} \, t + \bar{z} \frac{g^2}{4 x} \right)
\right]^\prime
\nonumber\\
&=&\!\!\!
i \theta_\pm
-
\int_{- 1}^1 \frac{dt}{\pi} \,
\frac{\ln Q \left( \pm \ft{i}{2} - g t \right)}{\sqrt{1 - t^2}}
\frac{\sqrt{u^2 - g^2}}{u + g t}
\, .
\ea
Here we integrated by parts in the second line in order to separate the components $\theta_\pm$
of the spin-chain quasimomentum $\theta = \theta_+ - \theta_-$,
\be
i \theta_\pm = \int_{- 1}^1 \frac{dt}{\pi} \,
\frac{\ln Q \left( \pm \ft{i}{2} - g \, t \right)}{\sqrt{1 - t^2}}
\, .
\ee
Notice that $\theta$ reduces to the one-loop expression \re{LOenergyANDquasimom} upon setting
$g = 0$. While the condition $t (u_n) = 0$ yields the all-order Bethe Ansatz equations suggested
in Ref.\ \cite{BeiSta05}.

The conjectured multi-loop Baxter equation \re{AllOrderBaxterEq} with \re{AllLoopSigma} and the known
pattern of renormalization of the conformal spin in field theories can be used to determine the
all-loop analytic expression for the anomalous dimensions in terms of the Baxter function. To this
end, recall that the conformal spin of Wilson operators $J^{{\scriptscriptstyle (0)}} = N + \ft12
L$ defining the quadratic Casimir $q_2^{{\scriptscriptstyle (0)}} = - J^{{\scriptscriptstyle (0)}}
(J^{{\scriptscriptstyle (0)}} - 1) - \ft14 L$ gets additive renormalization by the anomalous dimensions
$\gamma(g)$ of composite Wilson operators at higher orders in coupling, $J^{{\scriptscriptstyle (0)}}
\to J = N + \ft12 L + \ft12 \gamma(g)$. This conclusion arises from considerations of conformal Ward
identities for Green functions with conformal operator insertion \cite{BelMul98,BelKorMul05}. Then a
short inspection of the Baxter equation \re{AllOrderBaxterEq} with the dressing factors $\sigma_\pm$
in the form \re{DressingInfiniteSeries} demonstrates that the first term in the series of $\sigma_\pm
(x) = i \gamma_\pm (g)/x + \dots$ induces the shift of the conformal spin,
\be
J^{{\scriptscriptstyle (0)}} = N + \ft12 L
\to
J = N + \ft12 L + \ft12 \left( \gamma_+ (g) - \gamma_- (g) \right)
\, .
\ee
Consequently, we may naturally identify the addendum with the anomalous dimensions of a multiplicatively
renormalizable composite operators, $\gamma (g) = \gamma_+ (g) - \gamma_- (g)$. Making use of the
explicit form of the dressing factors $\sigma_\pm$, we find the integral representation of $\gamma (g)$
in terms of the solution to the Baxter equation,
\be
\label{AllOrderEnergy}
\gamma (g) = i \frac{g^2}{\pi}
\int_{-1}^1 dt \, \sqrt{1 - t^2}
\left[
\ln Q \left( \ft{i}{2} - g \, t \right)
-
\ln Q \left( - \ft{i}{2} - g \, t \right)
\right]^\prime
\, .
\ee
The Taylor expansion shows that the lowest three orders in $g^2$ coincide with Eq.\ \re{ADs3loops}.

The Baxter equation \re{AllOrderBaxterEq} can be solved analytically order-by-order in coupling constant
for specific values of $L$ and $N$, e.g., for $L = 4$, $N = 2$ eigenvalue with zero quasimomentum reads%
\footnote{This anomalous dimension, when related to Berenstein-Maldacena-Nastase operators \cite{Bei03a},
agrees with previous one-, two- and five-loop analyses of Refs.\ \cite{BiaEdeRosSta02}, \cite{BeiKriSta03}
and \cite{BeiDipSta04}, respectively.},
\be
\label{L4N2ADs}
\gamma (g)
=
\ft{5 \pm \sqrt{5}}{2} g^2
-
\ft{17 \pm 5 \sqrt{5}}{8} g^4
+
\ft{585 \pm 207 \sqrt{5}}{160} g^6
-
\ft{5185 \pm 2039 \sqrt{5}}{640} g^8
+
\mathcal{O} (g^{10})
\, .
\ee
However, it has a limited range of applicability being asymptotic in character: it allows to find
the anomalous dimensions up to order $\mathcal{O}(g^{2n})$ only for operators of length $L \geq n$.
This restriction arises from the breaking of its polynomiality above a boundary value of $n$, i.e.,
for $L \leq n$. Analogous limitations apply to the Bethe Ansatz equations of Ref.\ \cite{BeiDipSta04}.
A generic dependence of $\gamma (g)$ on the parameters $L$ and $N$ is not known however and below we
will develop an asymptotic scheme to find it in the large spin limit.

\section{Asymptotic expansion}

The large$-N$ behavior of anomalous dimension is of special interest in its own right since it
governs the Sudakov asymptotics of scattering amplitudes \cite{Col89,Kor88}, and in light of
gauge/string duality, for it can be compared (at strong coupling) to energies of quasiclassical
strings \cite{GubKlePol03,FroTse03,Kru02,BelGorKor06,SakSat06}. Recall at first that the anomalous
dimensions of twist$-L$ operators occupy a band of width $L - 2$, with the upper and lower
boundaries scaling like \cite{BraDerMan98,BelGorKor03}
\be
\gamma_{\rm lower} (g) = 2 \Gamma_{\rm cusp} (g) \ln N
\, , \qquad
\gamma_{\rm upper} (g) = L \Gamma_{\rm cusp} (g) \ln N
\, ,
\ee
and the coefficient $\Gamma_{\rm cusp} (g)$ being the cusp anomalous dimension \cite{Pol80,KorRad87},
known to one- \cite{Pol80}, two- \cite{KorRad87,KotLip02,KotLip03,BelGorKor03} and three-loop orders
\cite{MocVerVog04,KotLipOniVel04,BerDixSmi05}. The minimal anomalous dimension $\gamma_{\rm lower}
(g)$ of high-twist operators develops the asymptotic behavior identical to the one of twist-two
operators \cite{BelGorKor06,EdeSta06}. Since the single-logarithmic regime is realized for $L {\rm e}^L
\ll N$ with $L, N \to \infty$ \cite{BelGorKor06}, this allows one to evade the limitation of the
asymptotic character of the Baxter equation and to derive an all-loop equation for the cusp anomaly
$\Gamma_{\rm cusp}$.

Notice that although we have to solve the problem with large quantum numbers, we cannot apply
traditional WKB expansion for $Q (u)$ (see, e.g., Ref.\ \cite{BraDerMan98}) since the latter is
valid for the spectral parameter which scales as $u \sim N^1$ while the energy is determined by
the Baxter function $Q (u)$ evaluated at the argument $u = \pm \ft{i}{2} - g t$ which behaves as
$u \sim N^0$. Therefore, we have to resort to other techniques. To this end, we will use in the
following the approach developed in Refs.\ \cite{DerKorMan03,BelGorKor06} for one-loop anomalous
dimensions and which, as we will see momentarily, is easily generalizable beyond leading order of
perturbation theory.

\subsection{One-loop Baxter equation}

Let us briefly review the formalism of Refs.\ \cite{DerKorMan03,BelGorKor06} applied to the one-loop
Baxter equation \re{OneLoopBaxterEq}. Though we are interested only in the lowest energy curve, at
the beginning we will be general enough to discuss subleading trajectories as well in order to point
out approximations which have to be imposed to separate the lowest anomalous dimension only. In the
regime in question, the conserved charges are large $q_k^{{\scriptscriptstyle (0)}} \sim N^k$ and,
therefore, the transfer matrix is large $|t^{{\scriptscriptstyle (0)}} (u)| \gg 1$. Introducing a new
function
\be
\label{PhiRatioLO}
\phi^{{\scriptscriptstyle (0)}} (u)
=
\frac{Q^{{\scriptscriptstyle (0)}} (u + i)}{Q^{{\scriptscriptstyle (0)}} (u)}
\, ,
\ee
we can rewrite the Baxter equation in the form
\be
u_+^L \phi^{{\scriptscriptstyle (0)}} (u)
+
\frac{u_-^L}{\phi^{{\scriptscriptstyle (0)}} (u - i)}
=
t^{{\scriptscriptstyle (0)}} (u)
\, .
\ee
The solution to it is based upon different scaling behavior of the right- and left-hand sides with
$N$. For the spectral parameter $u \sim N^0$, the solution is given by an infinite fraction. Keeping
the leading terms only we come to two difference equations
\be
\label{ReducedBE}
u_+^L Q_+^{{\scriptscriptstyle (0)}} (u + i)
=
t^{{\scriptscriptstyle (0)}} (u) Q_+^{{\scriptscriptstyle (0)}} (u)
\, , \qquad
u_-^L Q_-^{{\scriptscriptstyle (0)}} (u - i)
=
t^{{\scriptscriptstyle (0)}} (u) Q_-^{{\scriptscriptstyle (0)}} (u)
\, .
\ee
The additive corrections to their right-hand sides go as $\mathcal{O} (1/q_n^{{\scriptscriptstyle
(0)}})$, where $q_n^{{\scriptscriptstyle (0)}}$ is a conserved charge which scales with the
maximal power of $N$. For cyclically symmetric states $\theta = 0$, the asymptotic solution to
\re{OneLoopBaxterEq} reads
\be
Q^{{\scriptscriptstyle (0)}} (u)
=
Q_+^{{\scriptscriptstyle (0)}} (u) Q_-^{{\scriptscriptstyle (0)}} (- \ft{i}{2} )
+
Q_-^{{\scriptscriptstyle (0)}} (u) Q_+^{{\scriptscriptstyle (0)}} ( \ft{i}{2} )
\, ,
\ee
in terms of the solution to the two-term recursion relations \re{ReducedBE} written with the
help of the roots $\delta_k$ of the transfer matrix $t^{{\scriptscriptstyle (0)}} (u) = 2 (u
- \delta_1) (u - \delta_2) \dots (u - \delta_L)$ \cite{BelGorKor06},
\be
Q_\mp^{{\scriptscriptstyle (0)}} (u)
=
2^{ \pm i u} \prod_{k = 1}^L
\frac{\Gamma (\pm i u + i \delta_k)}{\Gamma (\pm i u + \ft{1}{2})}
\, .
\ee
Now recall that we are interested only in the trajectory with the lowest energy only. The latter
does not depend on the twist of the operator, i.e., it is $L-$independent. The reason for this
being that for the corresponding state only the quadratic Casimir $q_2^{{\scriptscriptstyle (0)}}$
is large while all other integrals of motion become anomalously small. For the roots of the transfer
matrix this is translated into the statement that just two roots $\delta_1 = \delta_L$ are much
larger than the rest of $\delta$'s which are negligible \cite{BelGorKor06}, yielding the relation
\be
\delta_1^2 \simeq - q_2^{{\scriptscriptstyle (0)}}/2
\, .
\ee
In this case the genus$-(L - 2)$ hyperelliptic Riemann surface parametrizing the magnet, with its
moduli determined by the conserved charges $q_k^{{\scriptscriptstyle (0)}}$, degenerates into a
sphere, i.e., the spectral curve of twist-two operators \cite{BelGorKor06}. This implies that all
zones but one of allowed classical motion in separated variables collapse into points. In this
limit the transfer matrix reduces to $t^{{\scriptscriptstyle (0)}} (u) \simeq u^L
\tau^{{\scriptscriptstyle (0)}} (u) = u^L (2 - N^2/u^2 )$ and the solutions to the recursion
relations \re{ReducedBE} becomes symmetric under the interchange $u \to - u$ and equal,
$Q_+^{{\scriptscriptstyle (0)}} (u) = Q_-^{{\scriptscriptstyle (0)}} (u)$. In the infinite-spin
limit, we then find that the leading behavior of the Baxter function is
\be
\left( i \ln Q_\pm^{{\scriptscriptstyle (0)}} (u) \right)^\prime
= \psi (- i u + i \delta_1) + \psi (- i u - i \delta_1) + {\dots}
\simeq 2 \ln N + \dots
\, ,
\ee
where in the last step we imposed the condition that the evaluation of the anomalous dimensions
\re{LOenergyANDquasimom} requires $u \sim N^0$ and thus it can be neglected compared to $N$. This
consideration immediately suggests that for the minimal-energy trajectory in the single-logarithmic
asymptotics the dressing factors $u_\pm^L$ in the left-hand side of Eq.\ \re{ReducedBE} are
irrelevant. Thus they can be reduced to $u_\pm^L \to u^L$ and cancelled with the factor extracted
from the transfer matrix $t^{{\scriptscriptstyle (0)}} (u)$, making the equation $L-$independent,
as expected. The latter is clearly seen in the quasiclassical approach when one assumes the
spectral parameter to scale with $N$, i.e., $u = N \hat{u}$ and $\hat{u} \sim 1$. We will use
the same argument below to write the all-loop Baxter equation for the lowest trajectory.

\subsection{Beyond one loop}

Let us find the equation for the minimal trajectory starting from the multi-loop Baxter equation
\re{AllOrderBaxterEq}. Again, we have to separate only terms which generate leading behavior in
the large-spin limit. The transfer matrix degenerates on the minimal trajectory to the one of
twist-two operators, i.e., $t (u) \simeq u^{L - 2} (2 u^2 + q_1 u + q_2)$. Notice however that
only $\mathcal{O} (g^0)$ contributions to the charges $q_{1,2} (g)$ can induce the leading effect
in the large$-N$ limit since the quantum corrections grow at most logarithmically with $N \to \infty$.
Therefore, we can replace $t (u) \simeq t^{{\scriptscriptstyle (0)}} (u)$ in the right-hand side of
\re{AllOrderBaxterEq}. Hence the reduced Baxter equation admits the form
\be
{\rm e}^{\sigma_+ (x_+)} Q (u + i)
+
{\rm e}^{\sigma_- (x_-)} Q (u - i)
=
\tau^{{\scriptscriptstyle (0)}} (u) Q (u)
\, .
\ee
Introducing again the ratio of the Baxter functions $Q$ analogous to Eq.\ \re{PhiRatioLO}, we can write
again two asymptotic equations for the two components of $Q$. However, since we are interested solely
in the lowest trajectory, both equations generate the same contributions to the anomalous dimension.
Therefore, we may consider only one of the resulting equations, e.g.,
\be
\label{InterBaxter}
{\rm e}^{\sigma_+ (x_+)} Q (u + i)
=
\tau^{{\scriptscriptstyle (0)}} (u) Q (u)
\, .
\ee
Next, introducing the one- and all-loop Hamilton-Jacobi functions,
\be
S^{{\scriptscriptstyle (0)}} (u) = \ln Q^{{\scriptscriptstyle (0)}} (u)
\, , \qquad
S (u) = \ln Q (u)
\, ,
\ee
Eq.\ \re{InterBaxter} can be rewritten by virtue of the one-loop degenerate Baxter equation \re{ReducedBE}
for the lowest trajectory as follows
\be
\label{CuspEquation}
S (u + i) - S^{{\scriptscriptstyle (0)}} (u + i)
-
S (u) + S^{{\scriptscriptstyle (0)}} (u) + \sigma_+ (x_+) = 2 \pi i m
\, .
\ee
Here $m$ displays the ambiguity in choosing the branch of the logarithm. Since the anomalous
dimension \re{AllOrderEnergy} is expressed in terms of the derivative of the Hamilton-Jacobi
function, it is instructive to differentiate both side of Eq.\ \re{CuspEquation} with respect
to $u$. Using the perturbative decomposition of the all-order Hamilton-Jacobi function
\be
S (u) = S^{{\scriptscriptstyle (0)}} (u) + g^2 S_h (u)
\, , \qquad \mbox{with} \qquad
S_h (u) = \sum_{n = 1}^\infty g^{2 (n - 1)} S^{(n)} (u)
\, ,
\ee
and rescaling $S_h^\prime$ by extracting its single logarithmic behavior
\be
i S_h^\prime (u)
=
\Sigma (u) \ln N
\, ,
\ee
we finally arrive at the equation for the cusp anomaly
\be
\label{CuspEquationSigma}
\Sigma (u + i) - \Sigma (u)
+
\frac{1}{\sqrt{u_+^2 - g^2}}
\int_{-1}^1 \frac{dt}{\pi} \frac{\sqrt{1 - t^2}}{u_+ + g t}
\left[
2 + g^2 \Sigma (\ft{i}{2} - g t)
\right]
= 0
\, .
\ee
The cusp anomalous dimension is then found in terms of $\Sigma$ making use of Eq.\ \re{AllOrderEnergy}
as
\be
\label{CuspBaxter}
\Gamma_{\rm cusp} (g)
=
g^2 + g^4 \int_{-1}^1 \frac{dt}{\pi} \, \sqrt{1 - t^2} \,
\Sigma \left( \ft{i}{2} - g \, t \right)
\, .
\ee
As we will demonstrate below, there exists yet another expression for the cusp anomalous dimension
in terms of the rescaled Hamilton-Jacobi function $\Sigma$ which leads to realization of an iterative
perturbative structure of $\Gamma_{\rm cusp}$ in gauge theory. Eqs.\ \re{CuspEquationSigma} and
\re{CuspBaxter} are the main results of this sections. If one shifts the spectral parameter as
$u \to u - \ft{i}{2}$, one immediately realizes that the first two terms give the imaginary part
of $\Sigma$ for real $u$. Then the use of a dispersion relation for the rescaled Hamilton-Jacobi
function in the last term allows us to bring the equation into the form of a Fredholm equation of
the second kind. Then the large$-x$ asymptotics of the solution to this integral equation yields the
cusp anomaly $2 x[u] \, \Im{\rm m} S_h (u + \ft{i}{2}) |_{x [u] \to \infty} = - \left[ \Gamma_{\rm
cusp} (g)/g^2 \right] \ln N$. However below we choose a slightly different route to solve Eq.\
\re{CuspEquationSigma} at weak coupling.

\section{Weak-coupling expansion}
\label{WeakCouling}

We will seek the solution to the cusp equation \re{CuspEquationSigma} in the form \cite{FadKor95,Kor95}
\be
\label{FormSolut}
\Sigma (u) = \int_0^1 d \omega \ \omega^{i u - 1} \ \bar{\omega}^{- i u - 1} \
\widehat{\Sigma} \left( \ln \frac{\omega}{\bar{\omega}} \right)
\, ,
\ee
with $\bar{\omega} = 1 - \omega$. This integral representation immediately diagonalizes the
difference terms. The change of variables to $p = \ln \omega /\bar{\omega}$ brings Eq.\ \re{FormSolut}
into the form of a Fourier transform. However, before we proceed with the above transformation of Eq.\
\re{CuspEquationSigma}, we will manipulate it at first. We notice that the first term in the infinite
series expansion in Chebyshev polynomials in Eq.\ \re{DressingInfiniteSeries} is determined by the
all-order anomalous dimension. Therefore, we can separate it from the kernel and rewrite the equation
for the cusp anomaly $\Gamma_{\rm cusp}$ in a form which immediately suggests yet another relation of
the Hamilton-Jacobi function to the cusp anomalous dimension. Performing these steps, we find
\be
\label{CuspEqFourier}
\sinh \left( \frac{p}{2} \right) \, \widehat{\Sigma} (p)
+
\frac{\Gamma_{\rm cusp} (g)}{g^3} \, J_1 (g p)
+
\frac{g p}{2} \int_0^\infty d p^\prime \frac{{\rm e}^{- p^\prime/2}}{p - p^\prime}
\mathbb{U} (g p, g p^\prime)
\widehat{\Sigma} (p^\prime)
= 0
\, ,
\ee
where the kernel $\mathbb{U}$ is expressed in terms of the Bessel functions,
\be
\mathbb{U} (p, p^\prime)
=
J_1 (p) \left[ J_0 (p^\prime) - \frac{2}{p^\prime} J_1 (p^\prime) \right]
-
J_1 (p^\prime) \left[ J_0 (p) - \frac{2}{p} J_1 (p) \right]
\, .
\ee
An examination of Eq.\ \re{CuspEqFourier} immediately suggests that the last term dies out for $p \to 0$
much faster than the first two, which scale linearly with $p$. Therefore, we deduce yet another
representation for $\Gamma_{\rm cusp}$ in terms of the solution $\widehat{\Sigma}$ to the cusp equation
\re{CuspEqFourier}, namely,
\be
\label{SigmaZeroCusp}
\widehat{\Sigma} (0) = - \frac{\Gamma_{\rm cusp} (g)}{g^2}
\, .
\ee
At the same time, we can use Eq.\ \re{CuspBaxter} for the anomalous dimension in terms of the
Hamilton-Jacobi function, such that we get
\be
\label{MixingOrders}
\widehat{\Sigma} (0)
=
- 1
- g \int_0^\infty \frac{d p}{p} \, {\rm e}^{- p/2}
J_1 (g p) \widehat{\Sigma} (p)
\, .
\ee
This expression clearly displays the mixing of orders and thus exhibits an iterative structure
of the perturbative series in coupling constant, i.e., the cusp anomaly at higher orders can be
determined in terms of $\Sigma (p)$ at lower orders. Combining Eqs.\ \re{CuspEqFourier} --
\re{MixingOrders} together we reproduce the cusp equation derived in Ref.\ \cite{EdeSta06}.

Finally, let us solve the cusp equation perturbatively. Writing the expansion in coupling constant as
\be
\label{PertExpanSigma}
\widehat{\Sigma} (p)
=
\frac{p/2}{\sinh p/2} \sum_{n = 0}^\infty g^{2 n}
\widehat{\Sigma}_n (p)
\, ,
\ee
where the prefactor is extracted for the latter convenience, and substituting it into the cusp equation
\re{CuspEqFourier}, we find for the few lowest order functions
\baa
\widehat{\Sigma}_0 (p)
\!\!\!&=&\!\!\!
- 1
\, , \\
\widehat{\Sigma}_1 (p)
\!\!\!&=&\!\!\!
\frac{\pi^2}{12} + \frac{1}{8} p^2
\, , \\
\widehat{\Sigma}_2 (p)
\!\!\!&=&\!\!\!
- \frac{11}{720} \pi^4
+ \frac{1}{8} \zeta (3) p
- \frac{\pi^2}{96} p^2
- \frac{1}{192} p^4
\, , \\
\widehat{\Sigma}_3 (p)
\!\!\!&=&\!\!\!
\frac{73 \pi^6}{20160}
-
\frac{\zeta (3)^2}{8}
-
\left(
\frac{5}{16} \zeta (5)
+
\frac{\pi^2}{96} \zeta (3)
\right)
p
+
\frac{\pi^4}{480} p^2
-
\frac{1}{96} \zeta (3) p^3
+
\frac{\pi^2 p^4}{2304}
+
\frac{p^6}{9216}
\, , \\
\dots \, .
\eaa
The $p$-independent term in these expressions determines the cusp anomaly according to Eq.\
\re{SigmaZeroCusp}. The lowest six orders of $\Gamma_{\rm cusp}$ read
\ba
\label{GammaCupsp}
\Gamma_{\rm cusp} (g)
\!\!\!&=&\!\!\!
g^2 - \frac{\pi^2}{12} g^4 + \frac{11 \pi^4}{720} g^6
\\
&-&\!\!\!
\left( \frac{73 \pi^6}{20160} - \frac{\zeta (3)^2}{8} \right) g^8
+
\left(
\frac{887 \pi ^8}{907200}
-
\frac{\pi^2}{48} \zeta (3)^2
-
\frac{5}{8} \zeta (3) \zeta (5)
\right) g^{10}
\nonumber\\
&-&\!\!\!
\left(
\frac{136883 \pi^{10}}{479001600}
-
\frac{\pi^4}{240} \zeta (3)^2
-
\frac{5 \pi^2}{48} \zeta (3) \zeta (5)
-
\frac{51}{64} \zeta (5)^2
-
\frac{105}{64} \zeta (3) \zeta (7)
\right) g^{12}
\nonumber\\
&+&\!\!\!
\bigg(
\frac{7680089 \pi^{12}}{87178291200}
-
\frac{47 \pi^6}{48384} \zeta (3)^2
+
\frac{\zeta (3)^4}{64}
-
\frac{41 \pi^4}{1920} \zeta (3) \zeta (5)
-
\frac{17 \pi^2}{128} \zeta (5)^2
-
\frac{35 \pi^2}{128} \zeta (3) \zeta (7)
\nonumber\\
&&
\qquad\qquad\qquad\qquad\qquad\qquad\qquad\qquad
-
\frac{273}{64} \zeta (5) \zeta (7)
-
\frac{147}{32} \zeta (3) \zeta (9)
\bigg)
g^{14}
+
\dots
\, . \nonumber
\ea
The two- and three-loop coefficients agree with Feynman diagram calculations of Refs.\
\cite{KotLip02,KotLip03,BelGorKor03} and \cite{MocVerVog04,KotLipOniVel04,BerDixSmi05},
respectively, and the rest with available predictions of Ref.\ \cite{EdeSta06}. The
calculation can be extended to few dozens of terms in the series \re{PertExpanSigma},
but the results are too cumbersome to display here.

\section{Outlook}

In this note we proposed a multi-loop asymptotic Baxter equation for anomalous dimensions of arbitrary
twist$-L$, spin$-N$ single-trace holomorphic Wilson operators in maximally supersymmetric Yang-Mills
theory. We developed an approach for the asymptotic solution of the resulting equation for large values
of spin $N$ and derived an all-order equation for the cusp anomaly which governs the Sudakov asymptotics
of anomalous dimensions. The problem with the asymptotic nature of the equation was overcome by
studying the lowest-energy trajectory which is insensitive to the twist of the operator in the single
logarithmic regime $L {\rm e}^L \ll N$, $L, N \to \infty$.

There are many questions which remain to be addressed. One has to constrain the amount of ambiguity
left in restoration of higher loop effects from the lowest few terms of perturbative series for the
dressing factors. The analysis of the strong-coupling expansion of $\Gamma_{\rm cusp}$ is of special
interest in light of available predictions for it from string theory \cite{GubKlePol03}. A preliminary
analysis reveals however that $g = \infty$ is an essential singularity of the cusp equation. Next, one
has to understand how to incorporate wrapping effects to the Baxter equation \re{AllOrderBaxterEq} and
to identify a putative microscopic spin chain standing behind it. An ultimate goal would be to generalize
the all-order Baxter equation to all sectors of $\mathcal{N} = 4$ super-Yang-Mills theory which is
conceivably described by a long-range graded magnet.

\section*{Note added}

Recently a new calculation was published of the four-loop cusp anomalous dimension using the unitarity
technique \cite{BerCzaDixKosSmi06}. Their numerical finding explicitly demonstrates that the prediction
\re{GammaCupsp} based on the Baxter equation \re{AllOrderBaxterEq} with the dressing factor
\re{DressingInfiniteSeries} is incorrect starting from four loops. In a companion paper \cite{BeiEdeSta06},
a modified form of the cusp equation was proposed which takes into account a non-trivial dressing factor
in Bethe equations of Ref.\ \cite{BeiSta05}.

Presently we use the result of Ref.\ \cite{BerCzaDixKosSmi06,BeiEdeSta06} in order to fix the form of
the four-loop correction to the Baxter equation \re{AllOrderBaxterEq} and find anomalous dimensions
of local Wilson operators. It was suggested \cite{BerCzaDixKosSmi06}, that to reconcile within
error bars the result of their numerical calculation with the one coming from the cusp equation,
the sign of the $\zeta^2 (3)$ in four-loop contribution of Eq.\ \re{GammaCupsp} has to be flipped.
This requires the following additive modification of the four-loop cusp anomaly \re{GammaCupsp},
\be
\Gamma_{\rm cusp} (g)
=
- \frac{\zeta (3)^2}{4} g^8
+ \mathcal{O} (g^{10})
\, .
\ee
In order to generate it from the cusp equation, one has to add the following term\footnote{In
the unnumbered equations above \re{CuspEqFourier}, it yields corrections to the right-hand side
the equations, i.e., $\widehat{\Sigma}_2 (p) = \dots + \ft12 \alpha \, p$, $\widehat{\Sigma}_3 (p)
= \dots + \ft12 \alpha \zeta (3) + \ft{1}{24} \alpha \, p (\pi^2 + p ) $.} to the left-hand side
of Eq.\ \re{CuspEqFourier}
\be
\dots
+ 2 \alpha g^2 J_2 (g p)
\left(
1
+
g \int_0^\infty d p^\prime \, {\rm e}^{- p^\prime/2} \,
\widehat{\Sigma} (p^\prime) \frac{J_1 (g p^\prime)}{p^\prime}
\right)
+
\mathcal{O} (g^3)
\, ,
\ee
in agreement with Ref.\ \cite{BeiSta05}. Here the favored value of the constant is
$\alpha = \ft12 \zeta (3)$ \cite{BerCzaDixKosSmi06,BeiEdeSta06}. This translates into
a modification of the integrand in Eq.\ \re{CuspEquationSigma},
\be
\frac{1}{u_+ + g t}
\to
\frac{1}{u_+ + g t} + \frac{i \alpha g^4}{x_+^2}
+
\mathcal{O} (g^6)
\, .
\ee
A simple analysis allows to unambiguously restore the correction term to the dressing factors
\re{DressingInfiniteSeries} of the Baxter equation \re{AllOrderBaxterEq}. Namely, the former
get shifted as
\be
\sigma_\pm (x) \to \sigma_\pm (x) + \Delta_\pm (x)
\, ,
\ee
with
\be
\Delta_\pm (x)
=
\mp \frac{i \alpha g^6}{2 x^2} \int_{-1}^1 \frac{dt}{\pi} \, \sqrt{1 - t^2} \,
\left[ \ln Q ( \pm \ft{i}{2} - g t) \right]^\prime
+
\mathcal{O} (g^7)
\, .
\ee
Taking into account this extra term, the anomalous dimensions of Wilson operators acquire
additional contributions. For instance, the four-loop term in Eq.\ \re{L4N2ADs} gets corrected
by
\be
\label{L4N2ADaddendum}
\gamma (g) = \dots - \alpha \frac{5 \pm \sqrt{5}}{8} g^8 + \mathcal{O} (g^{10})
\, .
\ee
This explicitly demonstrates that the attempt to rescue the principle of maximal transcendentality
\cite{KotLipOniVel04} in the cusp anomalous dimension with $\alpha = \ft12 \zeta (3)$ results in
breaking of the rational form of anomalous dimensions of local Wilson operators, i.e., they
acquire transcendental addenda \re{L4N2ADaddendum} in addition to rational terms \re{L4N2ADs}.
At the current state-of-the-art of higher loop calculations such terms are not ruled out yet.
Within Mueller's cut vertex technique \cite{Mue78}, the main sources of transcendental constants
in local anomalous dimensions comes from virtual self-energy and vertex corrections with rational
terms being generated by real cuts. The finiteness of maximally supersymmetric Yang-Mills theory,
especially transparent in the light-cone gauge where Ward identities imply equality of the
vanishing beta function with all field renormalization constants, seems to suggests the absence
of transcendental constants in local anomalous dimensions. This question deserves however a
thorough study.

\vspace{0.8cm}

\noindent This work was supported by the U.S.\ National Science Foundation under grant no.\ PHY-0456520.



\end{document}